\documentclass[]{paper}
\usepackage{graphicx}  
\usepackage[utf8]{inputenc}
\usepackage{cmap}
\usepackage{amsmath}
\usepackage{graphicx}
\usepackage{amsmath}
\usepackage[usenames]{color}
\usepackage{colortbl}
\usepackage[14pt]{extsizes}
\PassOptionsToPackage{english}{babel}
\usepackage[english]{babel}
\graphicspath{{imagesa/}}
\usepackage[left=2.0cm,right=1cm,top=2cm,bottom=2cm,bindingoffset=0cm]{geometry}
\title{Spectral characteristics  of the antiferromagnetic spin-1/2
	Heisenberg model on the square lattice in a magnetic field}
\author{P.S. Savchenkov$^{1,2}$, A.F. Barabanov$^{3}$\\{\small $^{1}$  National Research Nuclear University MEPhI, 115409 Moscow, Russia } \\ {\small $^{2}$  National Research Centre “Kurchatov Institute”, 123182 Moscow, Russia }\\{\small $^{3}$ Vereshchagin Institute of High Pressure Physics, Russian Academy of Sciences, Moscow, Russia}}

\begin{document}

\maketitle
\begin{abstract}
	We predict that spin-waves in an ordered square quantum antiferromagnet in a transverse magnetic field (h) may demonstrate three modes of spin excitations. Starting from the self-consistent rotation-invariant Green’s function method, a new mean-field theory is constructed for h$\ne$0. The method preserves the translational and the axial symmetries, and provides exact fulfillment of the single-site  constraint for each of the three modes. We examine the dynamical structure factors $S^{\alpha\alpha}(\mathbf{k},\omega ),\alpha=$ x, y, z.
	It is shown, that the introduction of h leads to the hybridization of two degenerate spin modes due to the appearance of a nondiagonal on $\alpha, \beta$ spin-spin Green's functions.  The comparison of the theory with  the exact diagonalization study and with results on inelastic neutron scattering experiments is discussed at T = 0. We discuss also the correspondence of the theory to the existing theories, which allow only two spin excitations modes for the total $S(\mathbf{k},\omega )$.

\end{abstract}

\section{Introduction}
The consideration of a two dimensional quantum antiferromagnet (2D-AFM) is usually based on various versions of the spin-$\frac{1}{2}$ Heisenberg square-lattice model.

This model continues to be studied \cite{Werth,vvTranq,vvMan,VVsch,sar,Mikh16,Bishop17,Bishop19}. It is used to describe superconducting cuprates and related compounds, including recently synthesized molecular 2D AFM\cite{xiao,woodward,kwon2019,valkov17}. 

The model demonstrates the strong influence of quantum fluctuations on the properties of spin excitations. The inclusion of the magnetic field h should enhance the role of fluctuations and rearrange the spin properties of the system.  This problem is often studied based on the simplest Hamiltonian (without introducing spin frustration and spin anisotropy).

\begin{equation}
\label{Ham}
\widehat{H}=\widehat{J}+\widehat{h}^{z}=\frac{1}{2}J\sum _{\mathbf{n,g}}%
\widehat{\mathbf{S}}_{\mathbf{n}}\widehat{\mathbf{S}}_{\mathbf{n+g}}-h\sum_{%
	\mathbf{n}}\widehat{S}_{\mathbf{n}}^{z}
\end{equation}
where $\widehat{\mathbf{S}}_{\mathbf{n}}$ - is the quantum spin-$\frac{1}{2}$ operator on the site \textbf{n}, \textbf{g} - the vectors of the nearest neighbours on a square lattice in the XY plane. J is the AFM exchange parameter. The magnetic field h is applied perpendicular to the XY-plane. 

Most analytical considerations are based on the introduction of Bose operators using the $ \frac{1}{S} $ expansion in the Holstein – Primakov  method \cite{holst-prim}, or using the Dyson – Maleev method \cite{daison}.  
Such a consideration leads to the existence of one branch of Bose excitations \cite{Jit-nikuni98} in the case of low fields (h<0.7h$_{sat}$, h$_{sat}=4J$  is the saturation field,  at which spins become fully aligned).
The second mode of Bose excitations in the \cite {Jit-Cher_PRL99}  occurs only at h>0.7h$_{sat}$ when the decay of spin waves becomes important.

However, in the exact diagonalization method (ED) \cite{lusher}, the total dynamic structure factor S(\textbf{k},$\omega$) has a three-peak structure even in low fields. 
This kind of S(\textbf{k},$\omega$) cannot be described by the theory \cite{Jit-nikuni98, Jit-Cher_PRL99}.

 Let us mention the consideration of the model within the Tyablikov decoupling approximation. The approximation  leads to the presence of two modes of spin excitations \cite{jensen}.
 
Recent inelastic neutron scattering experiments also demonstrate the presence of several peaks of spin excitations in low magnetic fields.
In particular,  spin excitations branches splitting was established for 2D AFM Ba$_2$MnGe$_2$O$_7$ \cite{matsuda} even by h=0.3h$_{sat}$.

Similar measurements were carried out for a new class of molecular 2D-AFM \cite{Tsyrulin} and quasi two dimensional AFM \cite{Hong}.
The presence of several branches of spin excitations is experimentally observed in low magnetic fields, as in \cite{matsuda}. Branch splitting at the point \textbf{k}=($\pi$,$\pi$) increases linearly with h.

This work aims to establish that upon the introduction of h $\ne$0, three branches of spin excitations can be observed in the system. This physical picture qualitatively corresponds to the ED results \cite{lusher}.

Our consideration is based on a spherically symmetric self-consistent approach (SSSA) \cite{Mikh16, Kondo,Shim,Star92,muller}. Three degenerate branches of spin excitations are realized in the SSSA at h = 0. With the introduction of h$\ne$0 (the transition from O (3) symmetry to O (2) symmetry), a self-consistent theory for retarded Green's functions (GF) is developed.
Translational symmetry is preserved. A distinctive feature of the approach is the fulfillment of the spin one site constraint condition for each of the three branches of spin excitations.

In the discussion, the obtained spectra, as well as dynamical structure factors, are compared with the results on the ED \cite{lusher} and \cite{Jit-nikuni98,Jit-Cher_PRL99} theory.
Note that the proposed theory allows one to consider the case of finite temperatures T$\ne$0.

\section{Green's functions at a finite magnetic field. }

Here, we discuss the retarded GF \cite{tyab,zub} $G_{\mathbf{nm}}^{\alpha \alpha }=\langle \widehat{S}_{\mathbf{n}}^{\alpha }|\widehat{S}_{\mathbf{m}}^{\alpha }\rangle _{z=\omega +i\eta, \eta \rightarrow +0 }$  and their Fourier transform $G_{\mathbf{k}}^{\alpha \alpha }(\omega)=\langle \widehat{S}_{\mathbf{k}}^{\alpha }|\widehat{S}_{-\mathbf{k}}^{\alpha }\rangle_{z}$, $\alpha = x, y, z$ to calculate the dynamic properties and the spectrum of spin excitations for the system with the Hamiltonian (\ref{Ham}). 

At h=0 the system   with O(3)-symmetry Hamiltonian (\ref{Ham}) is in states with an average spin value at the site $\langle \widehat{\mathbf{S}}_{\mathbf{n}}^{\alpha }\rangle =0$ at an arbitrarily low temperature. Due to spherical symmetry $G_{\mathbf{nm}}^{\alpha \alpha }$ is independent of $\alpha$. The spectrum of spin excitations is threefold degenerate \cite{Shim,afb-obzor}.

The presence of h in (\ref{Ham}) distinguishes the z direction. Therefore one should  consider $G_{\mathbf{nm}}^{xx}=G_{\mathbf{nm}}^{yy}\neq G_{\mathbf{nm}}^{zz}$.

Let us introduce the relations between the dynamical structure factor $S^{\alpha \beta }(\mathbf{k},\omega )$, the static structure factor $\ S^{\alpha \beta }(\mathbf{k})$, and spin-spin correlation functions $\langle \widehat{S}_{\mathbf{n+l}}^{\alpha }\widehat{S}_{\mathbf{n}}^{\beta }\rangle
=c_{l}^{\alpha \beta }$:

\begin{equation}
\label{DST}
S^{\alpha \beta }(\mathbf{k},\omega )=-\frac{1}{\pi }(m(\omega )+1) \mathrm{Im} G^{\alpha \beta }_{\mathbf{k}}(\omega );
\end{equation}
\begin{equation}
\label{SST}
 S^{\alpha \beta }(\mathbf{k})=\int_{-\infty }^{\infty }d\omega \ S^{\alpha \beta }(\mathbf{k},\omega );
\end{equation}
\begin{equation}
\label{COR}
c_{l}^{\alpha \beta }=\frac{1}{N}\sum_{\mathbf{k}}e^{i\mathbf{kl}}S^{\alpha\beta }(\mathbf{k}),
\end{equation}
where  \textbf{l}=\textbf{g},\textbf{d},\textbf{2g} -  vectors to first, second and third nearest neighbours, $m(\omega )$ is the Bose distribution function.

Further, for GF $G_{\mathbf{nm}}^{\alpha \alpha }$ we will require the conditions of constraint on each site, namely $c_{\mathbf{l=0}}^{\alpha \alpha }$ =$\frac{1}{4}$. This is equivalent to the normalization conditions for static structure factors:

\begin{equation}
\label{constraint}
N^{-1}\sum_{\mathbf{k}}S^{\alpha\alpha }(\mathbf{k})=\frac{1}{4}.
\end{equation}

We introduce the chain of equations for GF's $G_{\mathbf{nm}}^{zz }$,  $G_{\mathbf{nm}}^{xx(yy)}$ and  $G_{\mathbf{nm}}^{yx }$, in accordance with \cite{zub}. Our approach implies the exact consideration of one- and two-site excitation operators  that arise when commutating $\widehat{S}_{\mathbf{n}}^{\alpha }$ with the Hamiltonian (\ref{Ham}). The approximation is required to close the chain of equations. In our case, it will correspond to decoupling only the three-site excitation operators $\widehat{S}_{\mathbf{n}}^{\alpha }$ related to different sites. The decoupling will preserve translational and O(2) symmetries. In particular, this means that $c_{\mathbf{l\neq 0}}^{xy}=0$.

The first equation for the $G_{\mathbf{nm}}^{zz}$ is:
\begin{equation}
\label{zz1}
zG_{\mathbf{nm}}^{zz}=\langle \lbrack \widehat{S}_{\mathbf{n}}^{z};\hat{J}]|\widehat{S}_{\mathbf{m}}^{z}\rangle_{z}
\end{equation}
The approach needs to consider the second step - the differentiation of the right side of (\ref{zz1}).
Taking into account that the commutator $[[\widehat{S}_{\mathbf{n}}^{\alpha };\hat{J}];\hat{h}]=0$, we have the following exact expression for $G_{\mathbf{k}}^{zz}(\omega)$:
\begin{equation}
\label{zz2}
\begin{gathered}
z^{2}G_{\mathbf{k}}^{zz}(\omega)=-4J(1-\gamma _{\mathbf{k}})(c_{\mathbf{g}}^{xx}+c_{\mathbf{g}}^{yy})+2J^{2}(1-\gamma _{\mathbf{k}})G_{\mathbf{k}}^{zz}(\omega)+J^{2}\langle \widehat{\mathbf{D}}_{\mathbf{n}}^{z}|\widehat{S}_{\mathbf{m}}^{z}\rangle _{\mathbf{k},z}\\
\widehat{\mathbf{D}}_{\mathbf{n}}^{\alpha }=\sum_{\beta }\mathbf{\{}\sum_{\mathbf{g;a}}\overline{\delta }_{\mathbf{a;}\overline{\mathbf{			\mathbf{g}}}}\left[ (\widehat{S}_{\mathbf{n+g}}^{\beta }\widehat{S}_{\mathbf{n}}^{\beta }\widehat{S}_{\mathbf{n+g+a}}^{\alpha }-\widehat{S}_{\mathbf{n+g}}^{\alpha }\widehat{S}_{\mathbf{n}}^{\beta
}\widehat{S}_{\mathbf{n+g+a}}^{\beta })\right] \mathbf{-}\\-\sum_{\mathbf{g;a}}\overline{\delta }_{\mathbf{a;\mathbf{g}}}\left[ (\widehat{S}_{\mathbf{n}}^{\beta }\widehat{S}_{	\mathbf{n+g}}^{\beta }\widehat{S}_{\mathbf{n+a}}^{\alpha }-\widehat{S}_{\mathbf{n}}^{\alpha }\widehat{S}_{\mathbf{n+g}}^{\beta }\widehat{S}_{\mathbf{n+a}}^{\beta })\right] \mathbf{\}},
\end{gathered}
\end{equation}
where $\gamma_{\mathbf{k}}=\frac{1}{2}(\cos k_{x}+\cos k_{y})$,$\ \delta _{\mathbf{l;m}}$ - is  the Kronecker symbol, $\overline{\delta }_{\mathbf{l;m}}=1-\delta _{\mathbf{l;m}}$, $\mathbf{a(g)}$ -  vectors of nearest neighbours, $\overline{\textbf{g}}=-\textbf{g}$.

Note that $\sum_{\mathbf{n}}\widehat{\mathbf{D}}_{\mathbf{n}}^{z}=0$. This means that in the limit when \textbf{k}$\rightarrow$0 the exact equation (\ref{zz2}) is closed. 
The spin excitation mode $\omega_{\mathbf{k}}^{z}$ corresponding to the $G_{\mathbf{k}}^{zz}$ is gapless at $\textbf{k}=0$. Such a mode must be present at any accurate approach. 

The operator $\widehat{\mathbf{D}}_{\mathbf{n}}^{z}$ contains three-site terms at different sites and, unlike the right side of (\ref {zz1}), allows one to decouple (\ref {zz2}) in the site representation.

Such a decoupling corresponds the approximation of the following form:
\begin{equation}
\label{D_ras00}
\begin{gathered}
\overline{\delta }_{\mathbf{a;}\overline{\mathbf{\mathbf{g}}}}(\widehat{S}_{\mathbf{n+g}}^{x}\widehat{S}_{\mathbf{n}}^{x})\widehat{S}_{\mathbf{n+g+a}}^{z}\approx
\overline{\delta }_{\mathbf{a;}\overline{\mathbf{\mathbf{g}}}}\langle \widehat{S}_{\mathbf{n+g}}^{x}\widehat{S}_{\mathbf{n}}^{x}\rangle \widehat{S}_{\mathbf{n+g+a}}^{z}
 \end{gathered}
 \end{equation}
 Decoupling (\ref {D_ras00}) leads to the replacement in (\ref{zz2}) of the operator $\widehat{\mathbf{D}}_{\mathbf{n}}^{z}$ by $\widetilde{\mathbf{\widehat{\mathbf{D}}}}_n^z$, which has the form:
\begin{equation}
\label{Dras}
\begin{gathered}
\widetilde{\mathbf{\widehat{\mathbf{D}}}}_{\mathbf{n}}^{z}\mathbf{=2}\alpha
^{zz}\mathbf{\{}4(c_{2g}^{xx}+2c_{d}^{xx})\widehat{S}_{\mathbf{n}}^{z}-(c_{2g}^{xx}+2c_{d}^{xx})\Sigma _{\mathbf{g}}\widehat{S}_{\mathbf{n+g}}^{z}+\\+c_{g}^{xx}[(\sum _{\mathbf{g}}\widehat{S}_{\mathbf{n+2g}}^{z}+2\sum _{\mathbf{d}}\widehat{S}_{\mathbf{n+d}}^{z})-3\sum _{\mathbf{g}}\widehat{S}_{\mathbf{n+g}}^{z}]\mathbf{\}.\ }
\end{gathered}
\end{equation}
It is taken into account that $c_{g(d,2g)}^{xx}=c_{g(d,2g)}^{yy}$ due to axial symmetry.

 The simplest vertex correction $\alpha ^{zz}$ is introduced for decoupling (\ref{Dras}). The value of $\alpha ^{zz}$ is determined self-consistently from the requirement of the constraint $c_{0}^{zz}=\langle \widehat{S}_{\mathbf{n}}^{z}\widehat{S}_{\mathbf{n}}^{z}\rangle =\frac{1}{4}$.
 The decoupling (\ref{Dras}) preserves the exact condition $\omega _{\mathbf{k\rightarrow 0}}^{z}=0$.

We emphasize that the theory developed here uses unambiguous decoupling in the site representation which is performed after the second stage of differentiation of GF. The approach differs significantly from the recently presented mean-field approach \cite{Werth}. In  \cite{Werth} the averages are taken in the momentum representation at the level of the Hamiltonian.

As a result,  one can derive the following expression for the $G_{\mathbf{k}}^{zz}(\omega)$ after decoupling:
\begin{equation}
\begin{gathered}
\label{GF_zz}
G_{\mathbf{k}}^{zz}(\omega)=\frac{F_{\mathbf{k}}^{zz}}{z^{2}-(\omega _{\mathbf{k}}^{z})^{2}};\\
F_{\mathbf{k}}^{zz}=-4J(c_{g}^{yy}+c_{g}^{xx})(1-\gamma _{\mathbf{k}});\\
(\omega _{\mathbf{k}}^{z})^{2}=-32J^{2}\alpha ^{zz}c_{g}^{xx}(1-\gamma _{\mathbf{k}})(1+\gamma _{\mathbf{k}}+\rho ^{zz});\\
\rho ^{zz}=\frac{1+4\alpha^{zz}(c_{2g}^{xx}+2c_{d}^{xx}+12c_{g}^{xx})}{-16\alpha^{zz}c_{g}^{xx}}.
\end{gathered}
\end{equation}
The value of $\rho ^{zz}$ is directly connected with the gap value $\omega _{\mathbf{k}=(\pi ,\pi )}^{z}$.

$G_{\mathbf{k}}^{zz}$ does not explicitly depend on the magnetic field h. The field influence is manifested over the values of $c_{g,d,2g}^{xx}$  which are calculated self-consistently.

Let us now represent exact equations for the GF $G_{\mathbf{k}}^{xx}(\omega)$. The equations have the form:

\begin{equation}
\label{xx1}
zG_{\mathbf{k}}^{xx}(\omega)=\langle \lbrack \widehat{S}_{\mathbf{n}}^{x};\hat{J}]|\widehat{S}_{	\mathbf{m}}^{x}\rangle _{z,\mathbf{k}}+ihG_{\mathbf{k}}^{yx}(\omega)
\end{equation}
\begin{equation}
\label{xx2}
z^{2}G_{\mathbf{k}}^{xx}(\omega)=-4J\ast \lbrack c_{\mathbf{g}}^{zz}+c_{\mathbf{g}}^{yy}](1-\gamma _{\mathbf{k}})+2J^{2}(1-\gamma _{\mathbf{k}})G_{	\mathbf{k}}^{xx}(\omega)+J^{2}\langle \widehat{\mathbf{D}}_{\mathbf{n}}^{x}|\widehat{S}_{\mathbf{m}}^{x}\rangle _{z\mathbf{k}}+ihzG_{\mathbf{k}}^{yx}(\omega)
\end{equation}
The form of Eqs.(\ref{xx1}, \ref{xx2}) are close to Eqs.(\ref{zz1}, \ref{zz2}). But here the term with h and nondiagonal in $\alpha\beta$ indexes GF $G_{\mathbf{\textbf{k}}}^{yx}(\omega)$ arises for the first time. 

The exact equation for the $G_{\mathbf{\textbf{k}}}^{yx}(\omega)$ gives:
\begin{equation}
\label{yx2}
z^{2}G_{\mathbf{k}}^{yx}(\omega)=-izJ\left\langle \widehat{S}_{\mathbf{n}}^{z}\right\rangle +2J^{2}(1-\gamma _{\mathbf{k}})G_{\mathbf{k}}^{yx}(\omega)+J^{2}\langle \widehat{\mathbf{D}}_{\mathbf{n}}^{y}|\widehat{S}_{\mathbf{m}}^{x}\rangle _{z\mathbf{k}}-ihzG_{\mathbf{k}}^{xx}(\omega),
\end{equation}
where $\left\langle \widehat{S}_{\mathbf{n}}^{z}\right\rangle$ is the average value of the operator $\widehat{S}^{z}_{\textbf{n}}$ on the site \textbf{n}.  $\left\langle \widehat{S}_{\mathbf{n}}^{z}\right\rangle$ does not depend on \textbf{n}. 

In the absence of a magnetic field h $G_{\mathbf{k}}^{xx(yy)}(\omega)$=$G_{\mathbf{k}}^{zz}(\omega)$ and $G_{\mathbf{k}}^{yx}(\omega)=0$. The spectrum of the spin excitations has a solution with three degenerate modes.

After (\ref{D_ras00})-type decoupling  $\widehat{\mathbf{D}}_{\mathbf{n}}^{\alpha}\rightarrow \widetilde{\mathbf{\widehat{\mathbf{D}}}}_{\mathbf{n}}^{\alpha}$ in (\ref{xx2}, \ref{yx2}) the system of equations (\ref{xx2}-\ref{yx2}) is becomes closed for $G_{\mathbf{k}}^{xx}(\omega)$ and $G_{\mathbf{k}}^{yx}(\omega)$. Here and below we assume that $\alpha ^{xx}=\alpha ^{yy}$. 

The solution for $G_{\mathbf{k}}^{xx}(\omega)$ has the form:

\begin{equation}
\begin{gathered}
\label{GF_xx}
G_{\mathbf{k}}^{xx}(\omega)=\frac{F_{\mathbf{k}}^{xx}(z^{2}-(\omega _{\mathbf{k}}^{x})^{2})+z^{2}h\left\langle \widehat{S}_{\mathbf{n}}^{z}\right\rangle }{(z^{2}-(\omega _{\mathbf{k}}^{x})^{2})^{2}-z^{2}h^{2}};\ \\
F_{\mathbf{k}}^{xx}=-4J( c_{\mathbf{g}}^{zz}+c_{\mathbf{g}}^{yy})(1-\gamma _{\mathbf{k}})\\
(\omega _{\mathbf{k}}^{x})^{2}=-16J^{2}\alpha
^{xx}(c_{g}^{xx}+c_{g}^{zz})(1-\gamma _{\mathbf{k}})(1+\gamma _{\mathbf{k}}+\rho ^{xx});\\
\rho ^{xx}=\frac{1+2\alpha^{xx}(c_{2g}^{xx}+2c_{d}^{xx}+c_{2g}^{zz}+2c_{d}^{zz}+12c_{g}^{xx}+12c_{g}^{zz})}{-8\alpha ^{xx}(c_{g}^{xx}+c_{g}^{zz})}.
\end{gathered}
\end{equation}
Significantly, that the denominator of $G_{\mathbf{k}}^{xx}(\omega)$ has the form of two degenerate spin excitations with a frequency $\omega _{\mathbf{k}}^{x}$, which are hybridized due to the term  $z^{2}h^{2}$. As a result, the denominator of $G_{\mathbf{k}}^{xx}(\omega)$ (\ref{GF_xx}) describes two modes of excitations with:
\begin{equation}
\label{wpm}
\omega _{\mathbf{k}}^{\pm }=\sqrt{\frac{h^{2}}{4}+(\omega _{\mathbf{k}}^{x})^{2}}\pm\frac{h}{2}
\end{equation}

The expression for the  $G_{\mathbf{k}}^{xx}(\omega)$ (\ref{GF_xx}) leads to the following form of the dynamical structure factor (\ref{DST}):
\begin{equation}
\begin{gathered}
\label{DST_xx}
S^{xx}(\mathbf{k},\omega )=S^{+}(\mathbf{k},\omega )+S^{-}(\mathbf{k},\omega )\\
S^{+}(\mathbf{k},\omega )=I^{+ }(\mathbf{k})[(m(\omega _{\mathbf{k}}^{+})+1)\delta (\omega -\omega _{\mathbf{k}}^{+})+m(\omega _{\mathbf{k}}^{+}))\delta (\omega +\omega _{\mathbf{k}}^{+})]\\
S^{-}(\mathbf{k},\omega )=I^{- }(\mathbf{k})[(m(\omega _{\mathbf{k}}^{-})+1)\delta (\omega -\omega _{\mathbf{k}}^{-})+m(\omega _{\mathbf{k}}^{-}))\delta (\omega +\omega _{\mathbf{k}}^{-})]\\
\end{gathered}
\end{equation}
here $I^{+}(\mathbf{k)}$ and $I^{-}(\mathbf{k)}$ are intensities of spin excitations  with $\omega _{\mathbf{k}}^{+}$ and $\omega _{\mathbf{k}}^{-}$:
\begin{equation}
\begin{gathered}
\label{Int_pm}
I^{+}(\mathbf{k)}\ =\frac{F_{\mathbf{k}}^{xx}}{2\sqrt{\frac{h^{2}}{4}+(\omega _{\mathbf{k}}^{x})^{2}}}+\frac{h\left\langle \widehat{S}_{\mathbf{n}}^{z}\right\rangle }{8\sqrt{\frac{h^{2}}{4}+(\omega _{\mathbf{k}}^{x})^{2}}}+\frac{\left\langle \widehat{S}_{\mathbf{n}}^{z}\right\rangle }{4}\\
I^{-}(\mathbf{k)=}\frac{F_{\mathbf{k}}^{xx}}{2\sqrt{\frac{h^{2}}{4}+(\omega	_{\mathbf{k}}^{x})^{2}}}+\frac{h\left\langle \widehat{S}_{\mathbf{n}}^{z}\right\rangle
}{8\sqrt{\frac{h^{2}}{4}+(\omega _{\mathbf{k}}^{x})^{2}}}-\frac{\left\langle \widehat{S}_{\mathbf{n}}^{z}\right\rangle }{4}
\end{gathered}
\end{equation}
 It turns out that $G_{\mathbf{k}}^{yy}(\omega)=G_{\mathbf{k}}^{xx}(\omega)$ and $S^{yy}(\mathbf{k},\omega )=S^{xx}(\mathbf{k},\omega )$.

One can show that the explicit equation of $G_{\mathbf{kk}}^{xy}(\omega)$ (\ref{yx2}) satisfies the exact condition $c_{0}^{xy}=\langle \widehat{S}_{\mathbf{n}}^{y}\widehat{S}_{\mathbf{n}}^{x}\rangle =\frac{1}{N}\sum_{\mathbf{k}}S^{yx}(\mathbf{k})=\frac{i}{2}\langle \widehat{S}_{\mathbf{n}}^{z}\rangle$ as well as $c_{\mathbf{l}\neq 0}^{xy}=0$. 

Six spin correlation functions $c_{g,d,2g}^{zz},c_{g,d,2g}^{xx}$ are calculated self-consistently using the equations for $G_{\mathbf{k}}^{zz}(\omega)$  (\ref{GF_zz}) and $G_{\mathbf{k}}^{xx}(\omega)$  (\ref{GF_xx}). Vertex corrections $\alpha ^{zz} $ and $\alpha ^{xx} $  are determined self-consistently from the constraint condition (\ref{constraint}).

The present approach allows us to consider both finite temperatures T $> 0 $ and the case T=0, which will be considered below. At T=0 one should introduce Bose condensation part \cite{Shim} of the spin correlation functions (\ref{COR}).

The average   $\left\langle \widehat{S}_{\mathbf{n}}^{z}\right\rangle$ is taken as  $\left\langle \widehat{S}_{\mathbf{n}}^{z}\right\rangle =\chi^{zz}(h)h$, where the magnetic susceptibility $\chi^{zz}(h)$ coincides with numerical calculations presented in \cite{chisl_lusher}. This will allow us to compare the results of the presented approach with the ED-results \cite{lusher}, which uses the same form of $\chi^{zz}(h)$ \cite{chisl_lusher}.

We restrict our attention to the low magnetic field case (h<2J), when the spin excitations instability is insignificant \cite{Mourigal}.

\section{Results and Discussions}
Let us discuss the form of the total dynamical structure factor $S(\mathbf{k},\omega )$ at T=0:
\begin{equation}
\begin{gathered}
\label{dst_o}
S(\mathbf{k},\omega )=S^{zz}(\mathbf{k},\omega )+S^{xx}(\mathbf{k},\omega)+S^{yy}(\mathbf{k},\omega )=S^{zz}(\mathbf{k},\omega )+2(S^{+}(\mathbf{k},\omega )+S^{-}(\mathbf{k},\omega ))
\end{gathered}
\end{equation}
It has the form:
\begin{equation}
\begin{gathered}
\label{dst_Tz}
S(\mathbf{k},\omega )=I^{z}(\mathbf{k)}\delta (\omega -\omega _{\mathbf{k}}^{z})+2I^{+}(\mathbf{k)}\delta (\omega -\omega _{\mathbf{k}}^{+})+2I^{-}(\mathbf{k)}\delta (\omega -\omega _{\mathbf{k}}^{-}) 
\end{gathered}
\end{equation}
here intensities $I^{+}(\mathbf{k)}$ and $I^{-}(\mathbf{k)}$ are defined by the expression (\ref{Int_pm}), $I^{z}(\mathbf{k)=}\frac{F_{\mathbf{k}}^{zz}}{\omega _{\mathbf{k}}^{z}}$.

Equation (\ref{dst_Tz}) gives that $S(\mathbf{k},\omega )$ have a three-fold structure at h$\ne$0. Intensities of the peaks  $I^{+}(\mathbf{k)}$ and $I^{-}(\mathbf{k)}$ explicitly depend on h.

\begin{figure}
	\centering
	\includegraphics[width=0.63\linewidth]{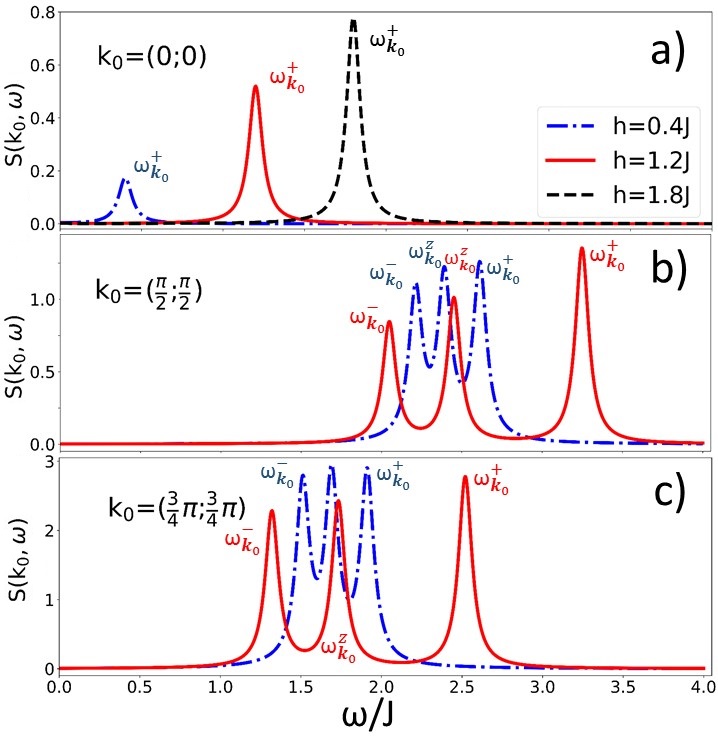}
	\caption[Short]{Dinamical structure factor  $S(\textbf{k},\omega)$ (\ref{dst_Tz}) for three points of the first Brillouin zone:
		
		 a) $\mathbf{k}_0=(0;0)$, b) $\mathbf{k}_0=(\frac{\pi }{2}; \frac{\pi }{2}),$ c) $\mathbf{k}_0=(\frac{3\pi }{4};\frac{3\pi }{4})$.
		 Blue (dash-dot) lines represent h=0.4J,  red (solid)  lines represent h=1.2J, black (dash) line represents h=1.8J.  
		 
		 $\omega _{\mathbf{k}_0}^{z},\omega _{\mathbf{k}_0}^{+},\omega _{\mathbf{k}_0}^{-}$ are related to intensities $I^{z}(\mathbf{k}_0)$, $I^{+}(\mathbf{k}_0)$ and $I^{-}(\mathbf{k}_0)$.}
	\label{fig:pic1}
\end{figure}

Fig. 1 represents $S(\mathbf{k},\omega )$ for three points of the first Brillouin zone: $\mathbf{k}_0=(0;0), \mathbf{k}_0=(\frac{\pi }{2};\frac{\pi }{2}),$ $\mathbf{k}_0=(\frac{3\pi }{4};\frac{3\pi }{4})$. An artifical broadening with Lorentzian type ($\delta=0.1J$) is introduced into the delta functions in (\ref{dst_Tz}).

Firstly, let us discuss $S(\mathbf{k}_0=0,\omega )$, see Fig.1a.  The peaks related to $\omega _{\mathbf{k}_0=0}^{z}$ and $\omega _{\mathbf{k}_0=0}^{-}$ are absent ($I^{-}(\mathbf{k_0=0})$=0 and $I^{z}(\mathbf{k_0=0)}$=0), $\omega _{\mathbf{k_0=0}}^{-}=\omega _{\mathbf{k_0=0}}^{z}=0$. Accordingly to (\ref{wpm}) $\omega _{\mathbf{k_0=0}}^{+}=h$. 
 
 $S(\mathbf{k}_0=0,\omega )$ was investigated by ED methods in \cite{lusher}. At Fig. 2 we compare the h dependence of $\omega _{\mathbf{k}_0=0}^{+}$ (\ref{wpm}) and $I^+(\textbf{k}=0)$(\ref{Int_pm}) with the ED results (see Fig.13 in \cite{lusher}). It may be seen that the presented approach reproduces the ED results: linear increase of  $\omega _{\mathbf{k}_0=0}^{+}$ with h and the increase of intensity.
 
Figs. 1b and 1c represent $S(\mathbf{k}_0,\omega )$ for typical points of Brillouin zone  corresponding to the essential  contribution to the constraint condition (\ref{constraint}). The splitting of the spin excitations bands into three peaks with frequencies $\omega _{\mathbf{k}_0}^{z},\omega _{\mathbf{k}_0}^{+},\omega _{\mathbf{k}_0}^{-}$ is observed. Such a splitting is similar to the ED results. For example, one can distinguish these modes at Fig.13(a,b) in \cite{lusher} ($\mathbf{k}_0=(\frac{\pi }{2};\frac{\pi }{2})$). According to (\ref{wpm}) splitting between modes $\omega _{\mathbf{k}}^{+}\ $ and $\omega _{\mathbf{k}}^{-}$ increases with h:  $\omega _{\mathbf{k}}^{+}\ -\omega _{\mathbf{k}}^{-}\ $=h. It turns that $\omega _{\mathbf{k}}^{z}$ satisfies the condition $\omega _{\mathbf{k}}^{-}<\omega _{\mathbf{k}}^{z}<\omega _{\mathbf{k}}^{+}$  for h$\neq 0, \mathbf{k\neq 0}$ . 

\begin{figure}
	\centering
	\includegraphics[width=0.58\linewidth]{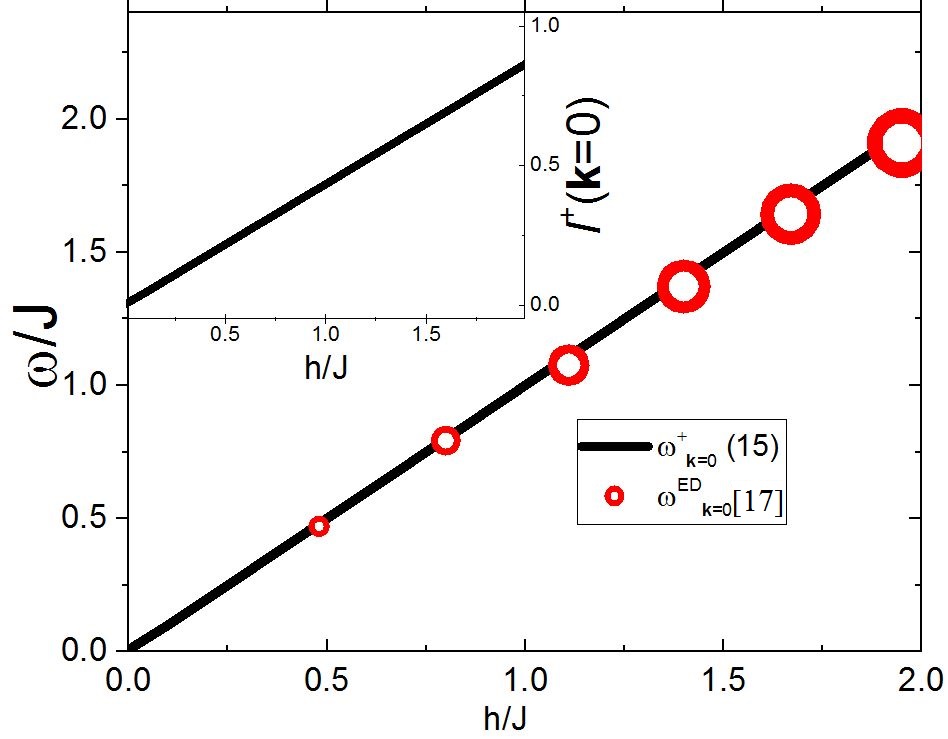}
	\caption{The field dependence of  $\omega _{\mathbf{k}_0=0}^{+}$ (line). Symbols show represent exact diagonalization results for $\omega _{\mathbf{k}_0=0}^{ED}$  \cite{lusher}. The area of the symbols is proportional to $I^{ED}(\textbf{k}=0)$. Inset: the field dependence of  $I^+(\textbf{k}=0)$(\ref{Int_pm}).}
	\label{fig:002}
\end{figure}

In a neighborhood of $\mathbf{k}_0=(\pi,\pi )$ three-peak structure may not be observed experimentally due to the proximity of frequencies $\omega _{\mathbf{k}}^{z}$  and $\omega _{\mathbf{k}}^{-}$. In this case, the peak structure corresponds to the results of neutron experiments for compounds: Cu(C$_4$H$_4$N$_2$)$_2$(ClO$_4$)$_2$ \cite{Tsyrulin}, C$_9$H$_18$N$_2$CuBr$_4$ \cite{Hong}.

Two modes of spin excitations are experimentally observed for Ba$_2$MnGe$_2$O$_7$ \cite{matsuda} at $\mathbf{k}=(\pi,\pi )$. The splitting between these modes is linear with h. The splitting is observed even in low fields coincidentally to S(\textbf{k},$\omega$) (\ref{dst_Tz}) with the condition of indistinguishability between $\omega _{\mathbf{k}}^{z}\ $ and $\omega _{\mathbf{k}}^{-}$.

Let us discuss the difference between our theory and approaches based on the introduction of Bose operators \cite{Jit-nikuni98,Jit-Cher_PRL99} in the framework of \cite{holst-prim,daison}.

 \begin{figure}
	\centering
	\includegraphics[width=0.85\linewidth]{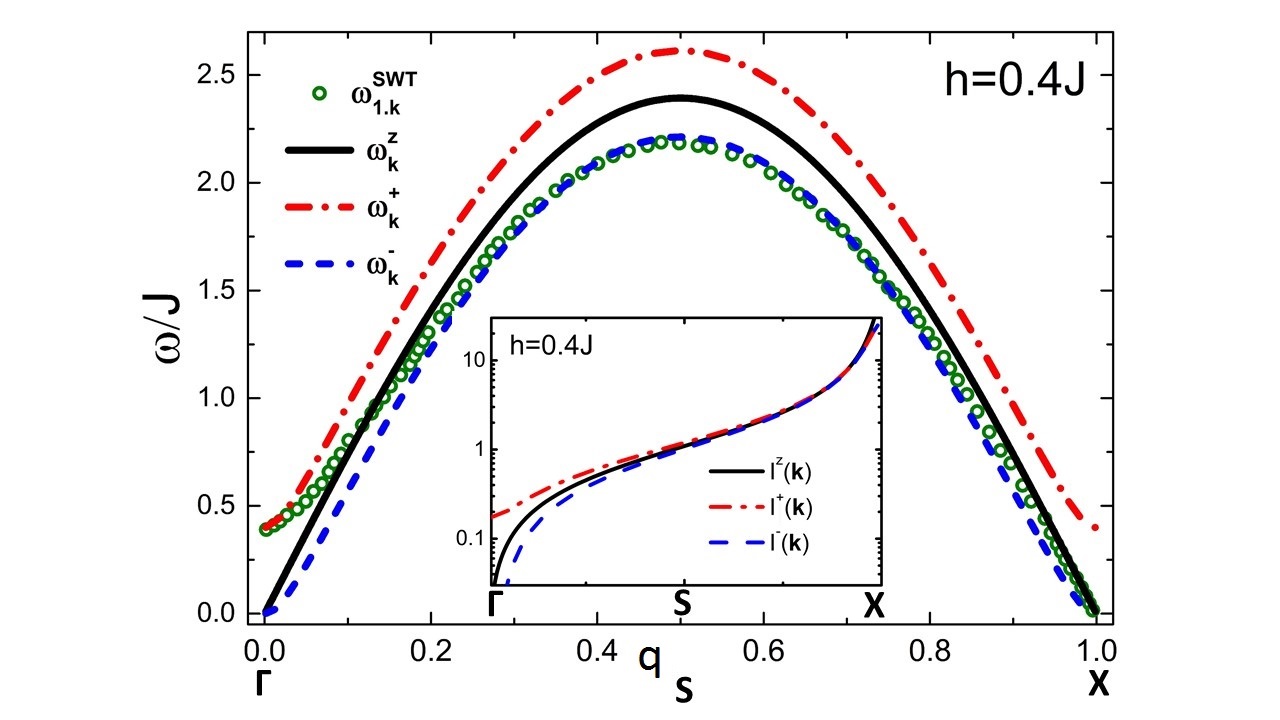}
	\caption{Magnon dispersion of three modes $\omega _{\mathbf{k}}^{+}$ (dash-dotted line), $\omega _{\mathbf{k}}^{z}$(solid line), $\omega _{\mathbf{k}}^{-}$(dashed line) for \textbf{k}=$\pi$(q,q). Symbols represent magnon dispersion of $ \omega_{1,\textbf{k}}^{SWT} $\cite{Jit-nikuni98}. Inset demonstrates the \textbf{k}-dependence of intensities: $I^+(\textbf{k})$(dash-dotted line),  $I^z(\textbf{k})$(solid line),  $I^-(\textbf{k})$(dashed line)  (\ref{dst_Tz}). } 
	\label{fig:jit}
\end{figure}

The expression for the dynamical structure factor of spin excitation S$^{SWT}$(\textbf{k},$\omega$) in \cite{Jit-Cher_PRL99} allows for the existence of two spin modes $ \omega_{1,\textbf{k}}^{SWT} $ and $ \omega_{2,\textbf{k}}^{SWT} =\omega_{1,\textbf{k}-\textbf{Q}}^{SWT}$ at low magnetic fields.
 These modes coincide along the magnetic Brillouin zone boundary. As a result, the structure of S$^{SWT}$(\textbf{k},$\omega$) is a single-peak along this boundary. In contrast, our theory (as well in \cite{lusher}) demonstrates a three-peak S(\textbf{k},$\omega$) structure (see Fig.1b). 
 
  It seems that the appearance of additional mode is possible in the approach [21]. In S$^{SWT}$(\textbf{k},$\omega$) cross-terms Green's functions were omitted as small numerically [16]. These terms can lead to hybridization splitting (15).
  
  Let us note, as mentioned in [16]  for the S=1/2 the quasiparticle weight redistribution is important. This redistribution is closely related to constraint condition (\ref{constraint}). The constraint condition is exactly fulfilled for $S^{xx(yy,zz)}(\mathbf{k},\omega )$. As to S$^{SWT}$(\textbf{k},$\omega$) this question remains open. 
  
To identify correspondence between theories  we represent the Fig. 3, where  $\omega _{\mathbf{k}}^{z},\omega _{\mathbf{k}}^{+},\omega _{\mathbf{k}}^{-}$   are demonstrated, as well as $ \omega_{1,\textbf{k}}^{SWT} $. The modes are given along the symmetric direction of the Brillouin zone $\Gamma$(($\mathbf{k}=(0 ,0 )$))-S($\mathbf{k=}(\frac{\pi }{2};\frac{\pi }{2})$)-X($\mathbf{k}=(\pi ,\pi )$) for h=0.4J. The intensities of modes (\ref{dst_Tz}) are shown at inset.

The mode $ \omega_{1,\textbf{k}}^{SWT}$ coincides  with the $\omega _{\mathbf{k}}^{+}$ in a small neighborhood of $\Gamma$.
 At $\Gamma$ only $I^{+}(\mathbf{k=0)}$ is finite. In this sense, $\omega _{\mathbf{k}}^{+}$  is the "leading"  mode. Both theories predict a linear dependence of these modes values on h.
 
 Values of $\omega _{\mathbf{k}}^{z},\omega _{\mathbf{k}}^{+},\omega _{\mathbf{k}}^{-}$  are well separated and  $ \omega_{1,\textbf{k}}^{SWT}$  coincides to $\omega _{\mathbf{k}}^{-}$ in a wide region of the phase space (neighborhood of the S-point). But the intensities $I^{-}(\mathbf{k})$, $I^{z}(\mathbf{k)}$, $I^{+}(\mathbf{k})$ are close to each other in this region.  

In a neighborhood of X $ \omega_{1,\textbf{k}}^{SWT}$ reproduces the "leading" $\omega _{\mathbf{k}}^{z}$ mode. Both of them are linearly dependent on momentum.

One can conclude that $ \omega_{1,\textbf{k}}^{SWT}$ coincides with the "leading" mode (if the "leading" mode can be distinguished).

Finally, we emphasize the main result of the present work. At h$\ne$0 a three-peak structure of the total dynamical structure factor S(\textbf{k},$\omega$) should be observed over a wide part of the Brillouin zone. This can be most clearly observed by inelastic neutron scattering at the boundary of the magnetic Brillouin zone. 

Let's pay attention to the rare earth compound YbB$_{12}$. The results of an experiment on inelastic neutron scattering\cite{nemkovsky} can be interpreted as the presence of two-dimensional AFM spin correlations in the compound. The peak corresponding to these correlations is significantly broadened with h at $\mathbf{k}=(\pi ,\pi )$. Perhaps this is evidence of a peak splitting, which is indistinguishable under the conditions of the experiment(h=10T, resolution equal to 0.5 meV). It seems important to study the structure of this peak at higher magnetic fields and with better resolution.

This work was supported by the Russian Foundation for Basic Research (project No.
19-02-00509).
This  work  was also supported by  the  Russian  Science  Foundation  (project  no.  18-12-00133).

The authors are grateful to P.A. Alekseev and  M.E. Zhitomirsky for useful discussions.


\begin{thebibliography}{10}
	\bibitem{Werth}
	Werth, A., Kopietz, P.,  Tsyplyatyev, O. (2018). Spin Hartree-Fock approach to studying quantum Heisenberg antiferromagnets in low dimensions. Physical Review B, 97(18), 180403.
	
	\bibitem{vvTranq}
	Tranquada, J. M. (2007). Neutron scattering studies of antiferromagnetic correlations in cuprates. In Handbook of High-Temperature Superconductivity (pp. 257-298). Springer, New York, NY.
	
	\bibitem{vvMan}
	Manousakis, E. (1991). The spin-1/2 Heisenberg antiferromagnet on a square lattice and its application to the cuprous oxides. Reviews of Modern Physics, 63(1), 1.
	
	\bibitem{VVsch}
	Schmidt, B., Thalmeier, P. (2017). Frustrated two dimensional quantum magnets. Physics Reports, 703, 1-59.
	
		\bibitem{sar}
	Sarıyer, O. S. (2019). Two-dimensional quantum-spin-1/2 XXZ magnet in zero magnetic field: Global thermodynamics from renormalisation group theory. Philosophical Magazine, 99(14), 1787-1824.
	
	\bibitem{Mikh16}
	Mikheyenkov, A. V., Shvartsberg, A. V., Valiulin, V. E., \& Barabanov, A. F. (2016). Thermodynamic properties of the 2D frustrated Heisenberg model for the entire J1–J2 circle. Journal of Magnetism and Magnetic Materials, 419, 131-139.
	

	
	\bibitem{Bishop17}
	Bishop, R. F., Li, P. H., Zinke, R., Darradi, R., Richter, J., Farnell, D. J. J., \& Schulenburg, J. (2017). The spin-half XXZ antiferromagnet on the square lattice revisited: A high-order coupled cluster treatment. Journal of Magnetism and Magnetic Materials, 428, 178-188.
	
	\bibitem{Bishop19}
	Bishop, R. F., Li, P. H. Y., Gotze, O.,  Richter, J. (2019). Frustrated spin-1 2 Heisenberg magnet on a square-lattice bilayer: High-order study of the quantum critical behavior of the J1-J2-J1 model. Physical Review B, 100(2), 024401.
	
	\bibitem{xiao}
	Xiao, F., Woodward, F. M., Landee, C. P., Turnbull, M. M., Mielke, C., Harrison, N., ... \& Pratt, F. L. (2009). Two-dimensional X Y behavior observed in quasi-two-dimensional quantum Heisenberg antiferromagnets. Physical Review B, 79(13), 134412.
	
	\bibitem{woodward}
	Woodward, F. M., Landee, C. P., Giantsidis, J., Turnbull, M. M., \& Richardson, C. (2001). Structure and magnetic properties of (5BAP) 2CuBr4: magneto-structural correlations of layered S= 1/2 Heisenberg antiferromagnets. Inorganica Chimica Acta, 324(1-2), 324-330.
	
	\bibitem{kwon2019}
	Kwon, S., Jeong, M., Kubus, M., Wehinger, B., Krämer, K. W., Rüegg, C., ... \& Lee, S. (2019). Field-induced anisotropy in the quasi-two-dimensional weakly anisotropic antiferromagnet [CuCl (pyz) 2] BF 4. Physical Review B, 99(21), 214403.
	
		\bibitem{valkov17}
	Val'kov, V. V., Dzebisashvili, D. M., Korovushkin, M. M., \& Barabanov, A. F. (2017). Stability of the d-wave pairing with respect to the intersite Coulomb repulsion in cuprate superconductors. Journal of Magnetism and Magnetic Materials, 440, 123-126.
	
	\bibitem{holst-prim}
	Holstein, T., \& Primakoff, H. (1940). Field dependence of the intrinsic domain magnetization of a ferromagnet. Physical Review, 58(12), 1098.
	
	\bibitem{daison}
	Dyson, F. J. (1956). General theory of spin-wave interactions. Physical review, 102(5), 1217.
	
	
	\bibitem{Jit-nikuni98}
	Zhitomirsky, M. E., \& Nikuni, T. (1997). Two-dimensional Heisenberg antiferromagnet in strong magnetic fields. Physica B: Condensed Matter, 241, 573-575.
	
	\bibitem{Jit-Cher_PRL99}
	Zhitomirsky, M. E., \& Chernyshev, A. L. (1999). Instability of antiferromagnetic magnons in strong fields. Physical review letters, 82(22), 4536.
	
		\bibitem{lusher}
	Lauscher, A., \& Lauchli, A. M. (2009). Exact diagonalization study of the antiferromagnetic spin-1/2 Heisenberg model on the square lattice in a magnetic field. Physical Review B, 79(19), 195102.
	
		\bibitem{jensen}
Jensen, P. J., Bennemann, K. H., Morr, D. K., \& Dreyssé, H. (2006). Two-dimensional Heisenberg antiferromagnet in a transverse field. Physical Review B, 73(14), 144405.

	\bibitem{tyab}
Tyablikov, S. V. The Methods in the Quantum Theory of Magnetism (Plenum Press, New York, 1967).
		\bibitem{zub}
	Zubarev, D. N. (1960). Double-time Green functions in statistical physics. Sov. Phys. Usp, 3(3), 320-345.
	
	\bibitem{matsuda}
	Masuda, T., Kitaoka, S., Takamizawa, S., Metoki, N., Kaneko, K., Rule, K. C., ... \& Nojiri, H. (2010). Instability of magnons in two-dimensional antiferromagnets at high magnetic fields. Physical Review B, 81(10), 100402.
	
	\bibitem{Tsyrulin}
	Tsyrulin, N., Pardini, T., Singh, R. R. P., Xiao, F., Link, P., Schneidewind, A., ... \& Kenzelmann, M. (2009). Quantum effects in a weakly frustrated S= 1/2 two-dimensional heisenberg antiferromagnet in an applied magnetic field. Physical review letters, 102(19), 197201.
	
	\bibitem{Hong}
Hong, T., Qiu, Y., Matsumoto, M., Tennant, D. A., Coester, K., Schmidt, K. P., ... \& Chernyshev, A. L. (2017). Field induced spontaneous quasiparticle decay and renormalization of quasiparticle dispersion in a quantum antiferromagnet. Nature communications, 8(1), 1-8.
	
	\bibitem{Kondo}
	Kondo, J., \& Yamaji, K. (1972). Green's-function formalism of the one-dimensional Heisenberg spin system. Progress of Theoretical Physics, 47(3), 807-818.
	
		
	\bibitem{Shim}
Shimahara, H., \& Takada, S. (1991). Green's Function Theory of the Two-DimensionalHeisenberg Model–Spin Wave in Short Range Order–. Journal of the Physical Society of Japan, 60(7), 2394-2405.
	
	\bibitem{Star92}
	AF, Barabanov, \& OA, Starykh (1992). Spherical symmetric spin wave theory of Heisenberg model. Journal of the Physical Society of Japan, 61(2), 704-708.
	
			\bibitem{muller}
	Muller, P., Lohmann, A., Richter, J., Menchyshyn, O.,  Derzhko, O. (2017). Thermodynamics of the pyrochlore Heisenberg ferromagnet with arbitrary spin S. Physical Review B, 96(17), 174419
	
	
		\bibitem{afb-obzor}
Barabanov, A. F., Mikheenkov, A. V., \& Shvartsberg, A. V. (2011). Frustrated quantum two-dimensional J1-J2-J3 antiferromagnet in a spherically symmetric self-consistent approach. Theoretical and Mathematical Physics, 168(3), 1192-1215.
	
		
	\bibitem{chisl_lusher}
	Richter, J., Schulenburg, J., \& Honecker, A. (2004). Quantum magnetism (Lecture Notes in Physics vol 645) ed Schollwöck U, Richter J, Farnell DJJ and Bishop RF.
		\bibitem{Mourigal}
	Mourigal, M., Zhitomirsky, M. E., \& Chernyshev, A. L. (2010). Field-induced decay dynamics in square-lattice antiferromagnets. Physical Review B, 82(14), 144402.
	

	\bibitem{nemkovsky}
	Nemkovski, K. S., Alekseev, P. A., Mignot, J. M., \& Ivanov, A. S. (2013). Resonant mode in rare-earth based strongly correlated semiconductors. Physics Procedia, 42, 18-24.
	

	


	



	
	
	
\end{thebibliography}
\end{document}